\documentclass[prd,aps,10pt,twocolumn,nofootinbib]{revtex4-2}

\usepackage{graphicx}
\usepackage{enumerate}
\usepackage{amsmath,amssymb,amsthm}
\usepackage{graphicx,caption}
\usepackage[paperwidth=210mm,paperheight=297mm,centering,hmargin=15mm,vmargin=20mm]{geometry}
\usepackage{float}
\usepackage{mathrsfs}
\usepackage{hyperref}
\usepackage{physics}
\usepackage[T1]{fontenc}
\usepackage{slashed}
\usepackage{xcolor}
\usepackage{xspace}

\usepackage{accents}

\renewcommand{\d}{\text{d}}
\newcommand{\e}{\text{e}}

\newcommand\ie{\mbox{\textit{i.\,e.}}\xspace}
\newcommand\eg{\mbox{e.\,g.}\xspace}

\date{\today}

\begin{document}

\author{Samuel Barroso Bellido}
\email{samuel.barroso-bellido@usz.edu.pl}
\affiliation{Institute of Physics, University of Szczecin, Wielkopolska 15, 70-451 Szczecin, Poland}

\author{Fabian Wagner}
\email{fabian.wagner@usz.edu.pl}
\affiliation{Institute of Physics, University of Szczecin, Wielkopolska 15, 70-451 Szczecin, Poland}

\title{A New Guest in the Third Quantized Multiverse}
% Alternative title: Taking third quantization seriously

\begin{abstract}

In the present article we point out a conceptual issue of the Third Quantization formalism for Canonical Quantum Gravity. Based on earlier results on interuniversal entanglement, the picture of noninteracting universe-antiuniverse pairs is incomplete. In particular, the variation of the entanglement entropy with the scale factor, understood as a measure of time, implies either an interaction between pairs of universes or universes with a heat bath, \ie an external field. We propose a simple solution to the problem by adding a new quantum field to the formalism. Correspondingly, particles of this field mediate interactions between universes -- depending on the model at tree level or one loop -- and may further provide a heat bath in the background.

\end{abstract}

\maketitle

\section{The Problem of Third Quantization of Canonical Quantum Gravity}

Canonical Quantum Gravity \cite{DeWitt,KieferQG} is directed at a description of the whole universe by means of a universal wave function $\Psi(g_{\mu\nu},\text{matter})$\cite{HH,VilenkinWF}. This wave function of the universe is considered to be a complex scalar in most of the literature except for some works in which the authors endow it with spinorial properties (see e.g. \cite{SpinorWF}) or define it over different fields (in the sense of Mathematics) like quarternions \cite{Quater}. Hence, as usual, we will consider it to be a complex scalar function.

Fundamentally, the wave function of the universe, living in the infinite-dimensional Hilbert space, the so-called superspace, satisfies the so-called Wheeler-DeWitt (WDW) equation whose general solution, however, is unknown. Therefore, by analogy with the case of classical General Relativity, it is useful to reduce the complexity of the problem by imposing symmetries. If we, for example, consider a FLRW universe with one degree of freedom representing the matter content, specifically a homogeneous scalar field $\phi$, the superspace is reduced to a two-dimensional space, the minisuperspace, whose variables are the said scalar field $\phi$ and the scale factor $a$, indicative for the expansion of the universe. For a more practical purpose, it is convenient to use a parameterization of the scale factor as $\alpha=\ln(a)$. In this case the WDW equation can be expressed as \cite{KieferQG,SBB}
\begin{equation}\label{WdW}
    \left[\pdv[2]{}{\alpha}-\pdv[2]{}{\phi}+m^2(\alpha,\phi)\right]\Psi(\alpha,\phi)=0,
\end{equation}
where the third term reads 
\begin{equation}\label{Mass}
m^2(\alpha,\phi)=\e^{6\alpha}\left[\frac{\Lambda}{3}+2V(\phi)\right]-\e^{4\alpha}K,
\end{equation}
in a rather general FLRW universe with curvature index $K$, a scalar field $\phi$ with potential $V(\phi)$ and a cosmological constant $\Lambda$, and the factor ordering has been chosen such that the momentum associated to the scale factor becomes 
\begin{equation}
    p_a^2=-\frac{1}{a}\pdv{}{a}\left(a\pdv{}{a}\right).
\end{equation}
Hence, the kinetic part of Eq. \eqref{WdW} can be written as the covariant d'Alembertian $\partial_{\mu}(\sqrt{g}g^{\mu\nu}\partial_{\nu})/\sqrt{g}$ on a two-dimensional Minkowski background. 
Since Eq. (\ref{WdW}) is very similar to a Klein-Gordon (KG) equation, given the hyperbolic nature of the differential operator, it is customary, but not necessary, to recognize $\alpha$ as a time variable, $\phi$ as a spatial variable and $m^2(\alpha,\phi)$ as a mass term. Some of the downsides of the resulting theory are discussed \eg in Refs. \cite{Prob1,Prob2} and will not be explicitly dealt with in the present work.

The formalism of Third Quantization \cite{3Q1,3Q2,3Q3,Salva3QReview} goes one step beyond in an attempt to construct a field theory describing the wave function of the universe $\Psi(\alpha,\phi)$ as a complex scalar field and the WDW equation (\ref{WdW}) as its equation of motion obtained through varying with respect to $\Psi^{\dagger}$
\begin{equation}
    \pdv{}{\alpha}\left(\frac{\delta\mathcal{L}}{\delta(\partial_{\alpha}\Psi^{\dagger})}\right)+
    \pdv{}{\phi}\left(\frac{\delta\mathcal{L}}{\delta(\partial_{\phi}\Psi^{\dagger})}\right)
    -\frac{\delta\mathcal{L}}{\delta \Psi^{\dagger}}=0,
\end{equation}
derived from the third quantized action
\begin{gather}\label{Action}
S=\int\d\alpha\d\phi\left[-\Psi^{\dagger}\square\Psi+m^2(\alpha,\phi)\Psi^{\dagger}\Psi\right],
\end{gather} 
where $\square=-\partial_{\alpha}^2+\partial_{\phi}^2$. Accordingly, the field $\Psi(\alpha,\phi)$ is treated as the representation of a universe-like particle, while $\Psi^{\dagger}$ yields the corresponding antiparticle. As a result, the multiverse is the playground of a bunch of particles $\Psi(\alpha,\phi)$ and $\Psi^{\dagger}(\alpha,\phi)$ interacting with each other, exclusively. Annihilation and creation operators have to be defined for each universe and antiuniverse in order to simplify the treatment, selecting also the proper Fock space as in Quantum Field Theory.

Based on the scheme presented by Third Quantization, the entanglement entropy of a pair of universes created as in analogy with pair creation \cite{SchwingerEffect,Pedro} has been recently calculated for different models in Refs. \cite{Interuniversal,SBB,SBB2}. It is important to notice that the number of particles is constant and any vertex stems from the action (\ref{Action}), so once the pair is born, neither of the two universes can decay or feel any other kind of interaction. If the multiverse was just an empty space where universes were freely moving, the entanglement entropy should not change unless they interact with an environment or another particle, which none of those cases are considered here. It was found that the entanglement entropy is not constant, in general \cite{SBB2}, for any value of $\alpha$ and $\phi$, and therefore there should be a source of its variability.

It may be argued that the entanglement entropy is not relevant but the production of entropy of the system \cite{KieferQG,ProdEntropy}
\begin{equation}
    \sigma=\dv{S}{a}-\frac{1}{T}\frac{\delta Q}{\d a},
\end{equation}
which considers it as a composite system evolving adiabatically. However, it does not stand up to the comparison with QED. Universes must be interpreted as point particles travelling through the vacuum and as such lack internal structure. Therefore, the net production of entropy is trivially zero, while the entanglement entropy remains nonvanishing in this picture.

Where does this entanglement of the pair thus come from? They are complex scalar particles created at the same point on the minisuperspace. Described by the natural Hamiltonian, usually called diagonal Hamiltonian since is separable as $
H=H_1+H_2$,
particles are not entangled. However, this picture looks unnatural since the number of universes apparently varies as time goes by \cite{Spectrum1,Spectrum2,Spectrum3,Spectrum4},  \ie as the universes evolve. This problem can be sorted out by choosing a certain representation in which the number of universes, or equivalently the universal number operator, remains constant, called the invariant or Lewis-Riesenfeld representation \cite{Lewis,SalvaInv}. Then, the quantum evolution of a pair is stationary, and so is the property of it being a pair. The comparison with the diagonal representation makes apparent that the Hamiltonian contains a non-diagonal part which is responsible for the entanglement \cite{Pedro}. This leads to the immediate conclusion that there is a wrong interpretation of Third Quantization since it is impossible for a pair to get a varying entanglement entropy in a non-interacting multiverse.

\section{The Solution: A forgotten guest}
Daring a closer look at the reasoning, the analogy with the KG equation was drawn in a rather na\"ive way. The Lagrangian of a field theory does not contain a function $m^2(\alpha,\phi),$ which is not described by a field, or the vacuum expectation value of a field which indeed would be constant, as it happens with the Higgs mechanism \cite{Higgs1,Higgs2}. In other words, there is no isomorphism between the KG equation and the WDW equation (\ref{WdW}) since the masses of both of them are essentially different. One is a constant, making the theory analytical manageable, and the other is a function of the internal degrees of freedom of the universes. Therefore, considering the scale factor as a time variable, the mass term changes in time. Any dynamics in the direction of the resulting arrow of time should thus be considered with due care.  Regarding the pair of particles, what it is shocking is the variability of the entanglement entropy in time. Besides, it is interesting to remark that the action (\ref{Action}) is not bounded from below when $m^2(\alpha,\phi)$ in Eq. \eqref{Mass} is negative. Yet, this argument may not be as strong as it seems since $\abs{\Psi(\alpha,\phi)}^2$ is not related to any probabilistic interpretation as in Quantum Mechanics \cite{TiplerWF,DressWF,VilenkinWFInter}.

In fact, it is more natural to understand the nonconstant mass term as an interaction with an external field. In that vein and by analogy \eg with QED, we may thus introduce a new degree of freedom $\Omega$ accounting for the dynamics of $m^2(\alpha,\phi)$ in the WDW equation \eqref{WdW}, which acts as a source term. Taking a step back by regarding the most general superspace, this new field would be the cause of the dynamics of the term proportional to \cite{KieferQG,Salva3QReview}
\begin{equation}
    \sqrt{h}\left[-^{(3)}R+2\Lambda+16\pi\left[K_\text{s}+V\right]_{\text{matter}}\right]\supset \mathcal{H},
\end{equation}
contained into the most general WDW equation, where $\mathcal{H}$ is the Hamiltonian, $^{(3)}R$ is the Ricci scalar of the spacial part, $K_{\text{s}}$ is the kinetic part of the spatial coordinates of the matter content, and $V$ the matter potential.

There are many actions from which the WDW equation \eqref{WdW} derives as an equation of motion. For example, let us consider a real scalar field $\Omega,$ leading to the dynamics
\begin{multline}\label{ActionOmega3}
    S_{3Q}
    =\int\d\alpha\d\phi
    \left[
    -\Psi^{\dagger}\square\Psi
    -\frac12\Omega\square\Omega
    +\right.\\\left.+g_3\Omega\Psi^{\dagger}\Psi
    +V(\Omega^2)
    \right],
\end{multline}
where $V(\Omega^2)$ is the particular potential of the new field $\Omega$, $g_3$ is a coupling constant and we get the relation $g_3\Omega=m^2$. The field $\Omega$ is expected to be real since it is proportional to $m^2(\alpha,\phi)$, which can be any real, possibly negative, number. The new field thus has its own dynamics ruled by the equation of motion
\begin{equation}\label{DynOmega3}
    -\frac{1}{g_3}\left\{\left[\pdv[2]{}{\alpha}-\pdv[2]{}{\phi}\right]\Omega+\pdv{V(\Omega^2)}{\Omega}\right\}=\Psi^{\dagger}\Psi.
\end{equation}
From this equation of motion it seems that there is a way to control the unitarity of the wave function, since there is an explicit form of $\abs{\Psi(\alpha,\phi)}^2$ given in terms of $\Omega$, but such control is not perfect. For instance, let us consider the potential $V(\Omega^2)=M_{\Omega}\Omega^2/2$ and that the wave function of the universe fulfills $\abs{\Psi(\alpha,\phi)}^2=C$, at all points of the phase space $\{\alpha,\phi\}$, where $C$ is a positive constant. Eq. (\ref{DynOmega3}) is then
\begin{equation}\label{DynOmega3Unit}
    \left[\pdv[2]{}{\alpha}-\pdv[2]{}{\phi}\right]\Omega+M_{\Omega}\Omega
    =-g_3.
\end{equation}
Taking a look at Eq. (\ref{Mass}), the limit at early times shows a vanishing function for $m^2(\alpha,\phi)$ and its derivatives, hence $\Omega$ is also vanishing. It is not in agreement with what we obtain from Eq. (\ref{DynOmega3Unit}) since $\Omega$ cannot be vanishing.
% Therefore, there is no way to keep a constant $\abs{\Psi}^2$ at least with the potential we used. Even if we find a general potential which can control the unitarity of the wave function of the universe, it would mean that it has a probabilistic interpretation after all, and so the action (\ref{ActionOmega3}) is not bounded from below. \textcolor{red}{\bf [What unitarity? $\Psi$ is not a wave function...]}

\begin{figure}[H]
\centering
\includegraphics[width=0.45\textwidth]{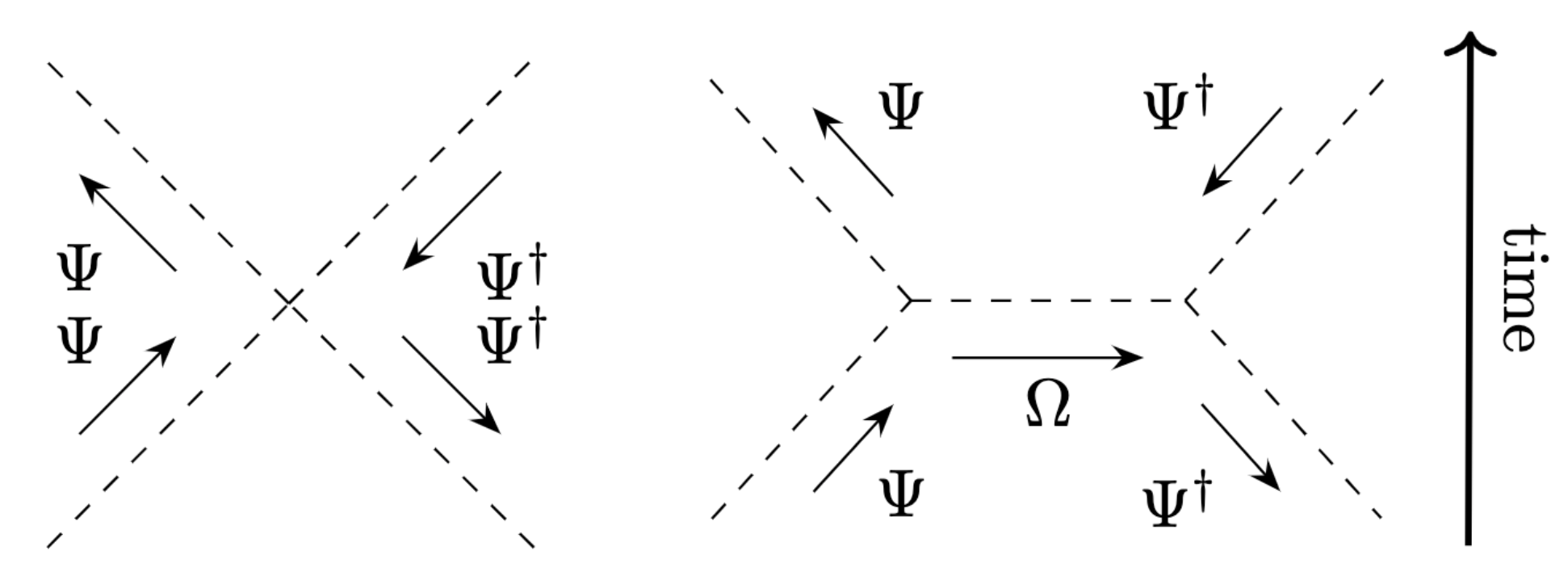}
\caption{Interaction between universes without the mediation by the $\Omega$-bosons (left) and with them (right) as derived from the action (\ref{ActionOmega3}), \ie the second example. Time's flow in this case implies a change in the variable $\alpha,$ \ie the scale factor.}
\label{Fig1}
\end{figure}

Alternatively, it is natural to consider a complex scalar field $\Omega$. In this case, the action reads
\begin{multline}\label{ActionOmega4}
    S_{3Q}
    =\int\d\alpha\d\phi
    \Big[
    -\Psi^{\dagger}\square\Psi
    -\Omega^{\dagger}\square\Omega
    +\\+g_4\Omega^{\dagger}\Omega\Psi^{\dagger}\Psi
    +V(\abs{\Omega}^2)
    \Big],
\end{multline}
from where we recognize $m^2=g_4\Omega^{\dagger}\Omega>0$.
It is important to notice that here the action is bounded from below, since $m^2(\alpha,\phi)>0$, which was our weak condition to get a bounded action \eqref{Action}, leading to a simplification in the physical interpretation. This scenario constraints all universes which do not fulfill the condition. In theory, there should not be any problem to describe the actual universe since the mass term in Eq. \eqref{WdW} $m^2(\alpha,\phi)$ is always positive during standard $\Lambda$CDM evolution \cite{WeinbergCosmo}. The dynamics of the new field $\Omega^{\dagger}$ is now governed by the equation of motion
\begin{equation}
\left[\pdv[2]{}{\alpha}-\pdv[2]{}{\phi}+g_4\Psi^{\dagger}\Psi\right]\Omega=-\pdv{V(\abs{\Omega}^2)}{\Omega^{\dagger}},
\end{equation}
while $\Omega$ obeys the complex conjugated equation. Both equations for the new particle are similar to a WDW equation \eqref{WdW}, and coincide when $V(\abs{\Omega}^2)$ vanishes. Whether self-interactions of the $\Omega$ field exist or not is impossible to say, and that is why we cannot conclude that it is a universe-like field like $\Psi$. Even so, it intentionally preserves the $U(1)$ symmetry found in the action \eqref{Action} such that the total electric charge and the number of particles continue to be conserved. This example is clearly more natural than the first one. However, in contrast with the latter model, it does not lead to a three-leg vertex. Correspondingly, the extensively-studied picture of parent universes giving birth to baby universes \cite{Baby2} is rendered impossible unless exotic ingredients like axion-like particles are included \cite{Baby1}.

\begin{figure}[H]
\centering
\includegraphics[width=0.45\textwidth]{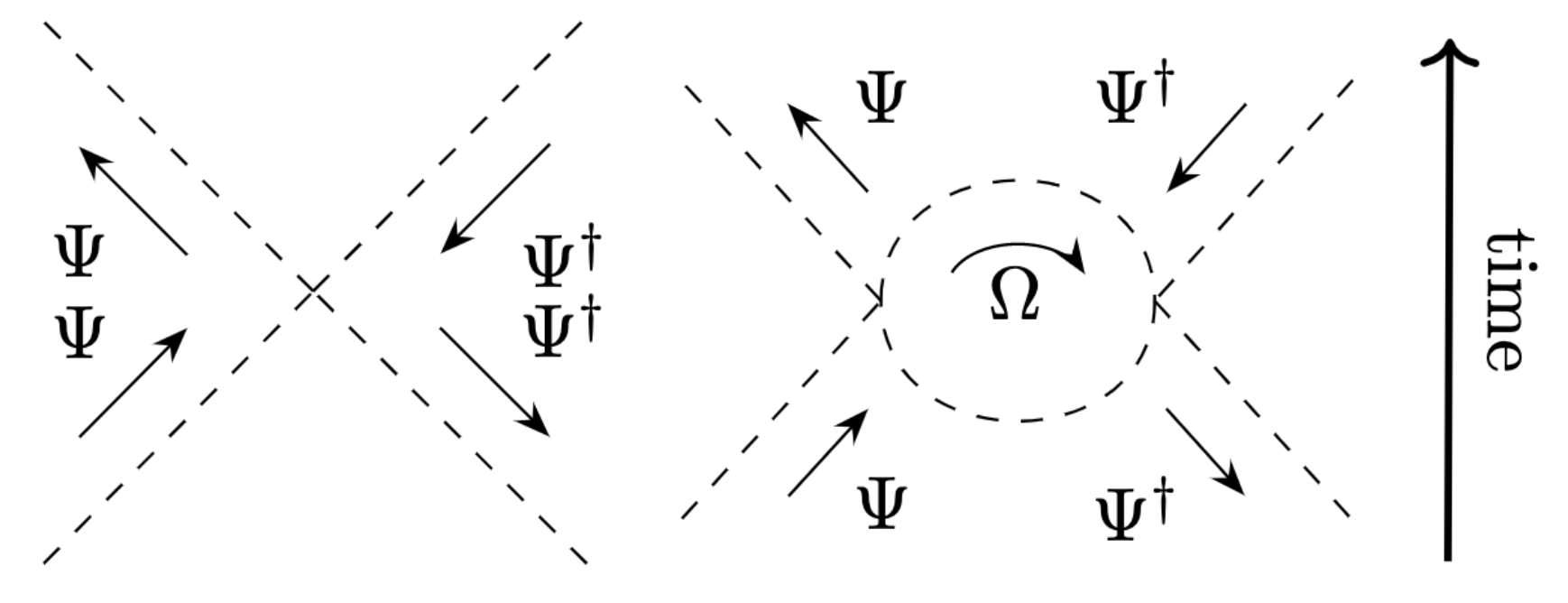}
\caption{Interaction between universes without the mediation by the $\Omega$-bosons (left) and with them (right) as derived from the action (\ref{ActionOmega4}), \ie the second example. Time's flow in this case implies a change in the variable $\alpha,$ \ie the scale factor.}
\label{Fig2}
\end{figure}

Nonetheless, most of the work done on Third Quantization is not spoiled since the WDW equation (\ref{WdW}) may be understood as an effective description of the underlying field theory we propose. Note that due to the "coordinate" dependence of the mass in the third quantized action \eqref{Action}, a nonperturbative solution of the path integral was not achievable in the first place, seemingly resulting in a hidden Fermi-like interaction provided on the left of Figures \ref{Fig1} and \ref{Fig2}. These diagrams are completed by the introduction of the real scalar and the complex scalar mediators to the right of Figures \eqref{Fig1} and \eqref{Fig2}, respectively. For instance, the entanglement entropy of the pair of universes calculated in Refs. \cite{Interuniversal,SBB,SBB2} remains unchanged for all models. However, the  spectra of the infinite universes created by dynamical variables in minisuperspace \cite{Spectrum1,Spectrum2,Spectrum3,Spectrum4} square well with this new interpretation because the background is flat, while an external field is held accountable for this variation. The resulting field theory on flat spacetime, including a source term, is thus equivalent to the canonical Third Quantization formalism on a dynamical, \ie expanding, minisuperspace. The only difference lies in the interactions between universes and antiuniverses by virtue of a new field. Hence, the concept of Third Quantization is carried through all the way inasmuch as every field is quantized instead of treating some of them as classical background.

\section{Conclusion}

Recent work on interuniversal entanglement consistently points towards a time evolution of corresponding entropy. This fact calls into question the concept of noninteracting pairs of universes without background. We have shown how the problem can be circumvented by introducing a new particle mediating interactions between universes and possibly providing a heat bath for decoherence. The resulting theory may be understood as a reformulation of Third Quantization such that the difference to standard free scalar field theory becomes manifest.

Although this formalism leads to the same predictions as standard Third Quantization, it makes apparent that different universes do in fact interact, which is the source for the variation of the interuniversal entanglement entropy. An interaction between universes, in turn, makes these observable at least in principle, thus possibly carrying the multiverse out of the realm of the metaphysical. It would be interesting to quantify such observables in future work.

\section{Acknowledgements}

S.B.B. would like to thanks L. M. Leza for the philosophical conversation that planted the seed for the present work. Both authors express their gratitude to S. Robles-Perez for his useful comments and corrections. This work has been supported by the Polish National Research and Development Center (NCBR) project UNIWERSYTET 2.0. STREFA KARIERY, POWR.03.05.00-00-Z064/17-00.

\bibliography{3QCorrection.bib}

\end{document}